\documentclass[intlimits,twoside,a4paper]{article}

\usepackage{amsmath,amssymb}
\usepackage{graphicx}
\usepackage{wrapfig}

\usepackage[T2A]{fontenc}
\usepackage[cp1251]{inputenc}
\usepackage[eqsecnum]{cmpj2}
\issue{2016}{19}{1}{13003}
\doinumber{10.5488/CMP.19.13003}

\title[Gyroidal nanoporous carbons~--- Adsorption and separation properties]%
{Gyroidal nanoporous carbons~--- Adsorption and separation properties explored using \\ computer simulations%
\thanks{This paper is dedicated to Prof. Stefan Soko\l{}owski on the occasion of his 65th birthday.}}
\author[S. Furmaniak \textsl{et al}.]{S. Furmaniak\refaddr{label1},
                            P.A. Gauden\refaddr{label1},
                            A.P. Terzyk\refaddr{label1},
                            P. Kowalczyk\refaddr{label2}}
\addresses{
\addr{label1} Faculty of Chemistry, Physicochemistry of Carbon Materials Research Group, \\ Nicolaus Copernicus University in Toru\'{n}, Gagarin St. 7, 87–100 Toru\'{n}, Poland
\addr{label2} School of Engineering and Information Technology, Murdoch University, \\ Murdoch,
Western Australia 6150, Australia
}

\date{Received November 13, 2015, in final form December 17, 2015}

\begin{document}

\maketitle

\begin{abstract}
Adsorption and separation properties of gyroidal nanoporous carbons (GNCs)~--- a new class of exotic nanocarbon materials are studied for the first time using hyper parallel tempering Monte Carlo Simulation technique. Porous structure of GNC models is evaluated by the method proposed by Bhattacharya and Gubbins. All the studied structures are strictly microporous. Next, mechanisms of Ar adsorption are described basing on the analysis of adsorption isotherms, enthalpy plots, the values of Henry’s constants, $\alpha_\textrm{s}$ and adsorption potential distribution plots. It is concluded that below pore diameters ca. 0.8~nm, primary micropore filling process dominates. For structures possessing larger micropores, primary and secondary micropore filling mechanism is observed. Finally, the separation properties of GNC toward CO$_2$/CH$_4$, CO$_2$/N$_2$, and CH$_4$/N$_2$ mixtures are discussed and compared with separation properties of Virtual Porous Carbon models. GNCs may be considered as potential adsorbents for gas mixture separation, having  separation efficiency similar or even higher than activated carbons with similar diameters of pores.

\pacs 05.10.Ln, 05.70.-a, 51.10.+y, 51.30.+i, 64.75.+g
\end{abstract}

\section{Introduction}

Adsorption of gas mixtures on solid surfaces has been attracting great interest for many years \cite{Ref01,Ref02,Ref03,Ref04,Ref05,Ref06,Ref07}. In the last decades there has been observed an increased interest to the development of new separation and purification techniques. It is evident that adsorption using various adsorbents is still a versatile tool for these purposes. On the other hand, the basic problem appearing in experimental studies is caused by difficulties in the synthesis of nanomaterials possessing desired properties. It is still not simple to control porosity and/or the chemical nature of the surface, and the both parameters at the same time. Moreover, experimental data on mixed systems are very limited, i.e., in the case of mixtures consisting of two volatile components the problem of the surface coverage determination has not been fully solved yet. Predicting adsorption behaviour of mixtures from pure component data is very important, from both the theoretical and practical viewpoints \cite{Ref08,Ref09,Ref10,Ref11,Ref12,Ref13,Ref14}. It is well known that the theoretical calculations provide additional opportunities for studies to better understand the separation processes. However, despite the intensive experimental and theoretical studies, our knowledge of the properties and the structure of mixed adsorbed layers is rather sparse, especially, on new generations of nanoporous adsorbents.

Computer simulation is an efficient method for resolving the above mentioned problems, since it is capable of modelling the processes of interest at the required level of detail in a controllable environment,  providing the necessary tools for establishing the connections between the observed phenomena and their molecular-level physical background. Prof. Soko\l{}owski’s (and co-workers) research topics of interest have also been concerned with the issue of mixtures using Monte Carlo simulations \cite{Ref15,Ref16,Ref17,Ref18,Ref19,Ref20,Ref21,Ref22}, Density Functional Theory \cite{Ref23,Ref24,Ref25,Ref26,Ref27,Ref28}, and Dissipative Particle Dynamics \cite{Ref29}. Their inspiring articles discuss, for example, the following problems: (i) adsorption from mixtures of monomers \cite{Ref15}, dimers \cite{Ref15}, the chain molecules \cite{Ref30,Ref31}, and even polymer mixtures \cite{Ref27}, (ii) adsorption from mixtures on homogeneous \cite{Ref25,Ref32} and heterogeneous surfaces \cite{Ref15,Ref33,Ref34}, (iii) layering transition, capillary condensation, wetting phenomena, and multilayer adsorption of binary ideal mixtures, systems exhibiting negative deviations from ideal mixing or positive one, binary mixture with partially miscible components, etc. \cite{Ref16,Ref17,Ref34,Ref35,Ref36,Ref37,Ref38}, (iv) interaction of charged chain particles and spherical counterions in contact with charged hard wall \cite{Ref31}, (v) analysis of the properties of two-dimensional symmetrical mixtures in an external field of square symmetry \cite{Ref39,Ref40}, (vi) demixing and freezing in two-dimensional symmetrical mixtures \cite{Ref21}, and (vii) the behaviour of mixed two component submonolayer films (Ar and Kr \cite{Ref41,Ref42} or Kr and Xe \cite{Ref22} on graphite). The majority of the analysed adsorbents have an ideal geometry of pores, for example a slit-like \cite{Ref25,Ref26,Ref32,Ref34,Ref43}. However, more complex models have also been studied, for example, pillared slit-like pores \cite{Ref28} and slit-like pores with walls decorated by tethered polymer brushes in the form of stripes \cite{Ref29}.

In the last decades, novel exotic porous carbon nanostructures (such as carbon nanotubes (CNTs), single-walled carbon nanohorns, graphene and graphitic nanoribbons, ordered porous carbons, wormlike nanotubes and graphitic nanofibers, stacked-cup carbon nanofibers, cubic carbon allotropes, carbon onions, carbyne networks, and others) have been projected to be among the most useful materials for selective adsorption and separation of fluid mixtures \cite{Ref44,Ref45,Ref46,Ref47,Ref48}. However, in the theoretical studies, different carbon adsorbents are studied, such as: carbon nanotubes \cite{Ref13,Ref49,Ref50,Ref51,Ref52,Ref53}, carbon nanohorns \cite{Ref13,Ref51,Ref54}, 2D and 3D ordered carbon networks \cite{Ref55}, hydrophobic virtual porous carbons (VPCs) \cite{Ref12,Ref14,Ref56,Ref57,Ref58,Ref59,Ref60,Ref61,Ref62}, oxidized VPCs \cite{ Ref12,Ref14,Ref60,Ref62,Ref63}, and triply periodic carbon minimal surfaces (Schwarz’s primitive and Schoen’s gyroid) \cite{Ref45,Ref59,Ref64,Ref65,Ref66,Ref67,Ref68,Ref69,Ref70}. Recently, scientists have paid attention to the next generation of porous carbon molecular sieves materials, i.e., crystalline exotic cubic carbon allotropes: cubic carbon polymorphs (CCPs) \cite{Ref45,Ref71,Ref72,Ref73}, diamond-like super structures of CNTs (super diamonds) \cite{Ref74}, diamond-like frameworks \cite{Ref75}, porous aromatic frameworks (PAFs) \cite{Ref76,Ref77}, diamond-like carbon frameworks (i.e., diamondynes, also named D-carbons) \cite{Ref78}, tetrahedral node diamondyne \cite{Ref79}, carbon allotropes proposed by Karfunkel and Dressler \cite{Ref45,Ref80}, compressed carbon nanotubes \cite{Ref45,Ref81}, sodalite-like nanostructures \cite{Ref45,Ref82}, folding of graphene slit-like pore walls \cite{Ref52,Ref83}, gyroidal nanoporous and mesoporous carbons (GNCs and GMCs, respectively) \cite{Ref84,Ref85,Ref86}.

One of the most interesting and promising adsorbent from the above mentioned  is GNC. In the current study, we consider nine different GNC structures having surface built in a way ensuring connection of each carbon atom with exactly three neighbours, similarly as ``schwarzites''. Nicola\"{i} et al. \cite{Ref84} confirmed that the curvature and the rigidity do not play a crucial role in the performance of GNC structures for ionic conduction. The major role, however, is played by the pore size and pore volume. Indeed, the larger the pore is, the larger is the ionic transport. Finally, the mentioned authors stated that GNC structures with tunable properties can be widely applied to water filtration or energy storage.

\section{Simulation details}

We used the structures of nine gyroidal nanocarbons (denoted as GNC-04, GNC-07, GNC-09, GNC-11, GNC-12, GNC-13, GNC-15, GNC-18 and GNC-21) published previously by Nicola\"{i} et al. \cite{Ref84} (see figure~1 in \cite{Ref84}). In the case of the first six systems, original boxes generated by Nicola\"{i} et al. \cite{Ref84} were multiplied (eightfold) in order to obtain box size at least two times greater than the cut-off distances used during simulations described below. The porosity of all the studied carbonaceous adsorbents was characterised by a geometrical method proposed by Bhattacharya and Gubbins (BG) \cite{Ref87}. The implementation of the method was described in detail elsewhere \cite{Ref88,Ref89}. The BG method provided histograms of pore sizes (effective diameters~--- ${d}_{\textrm{eff}}$). These data were also used to calculate the average sizes of pores accessible for Ar atoms (${d}_{\textrm{eff,acc,av}}$) \cite{Ref88}. In addition, the volume of pores accessible for Ar was determined using a combination of the BG method and Monte Carlo integration \cite{Ref88}.

Argon adsorption isotherms at its boiling point (i.e., ${T}=87$~K) on all the studied nanocarbons were simulated using the hyper parallel tempering Monte Carlo method (HPTMC) proposed by Yan and de Pablo \cite{Ref90}. The simulation scheme was the same as in previous work \cite{Ref88}. We considered 93 replicas corresponding to different relative pressure values (${p/p}_{\textrm{s}}$, where ${p}$ and ${p}_{\textrm{s}}$ are equilibrium and saturated Ar vapour pressure, respectively) in the range $1.0\times 10 ^{{-10}}-1.0$. Other details of the performed HPTMC simulations are described in \cite{Ref88}. The average numbers of Ar atoms in the simulation box were used to calculate the adsorption amount of Ar per unit of mass of the adsorbent (${a}$) \cite{Ref88}. The isosteric enthalpy of adsorption (${q}^{\textrm{st}}$) was also determined from the theory of fluctuations \cite{Ref88,Ref91} to reflect the energetics of the process.

In order to analyse the mechanism of Ar adsorption, we constructed high resolution $\alpha_{\textrm{s}}$-plots \cite{Ref92} based on simulated adsorption isotherms. We used Ar adsorption isotherm simulated in the ideal slit-like system composed of graphene sheets with effective pore width equal to 10~nm as the reference one. We also determined the values of Henry’s constant (${K}_{\textrm{H}}$) from the slope of the  linear part of adsorption isotherms in the low-pressure range \cite{Ref83}. Finally, adsorption potential distribution (APD) curves \cite{Ref93,Ref94,Ref95} were calculated. The APD curve is the first derivative of the so-called characteristic curve, presenting adsorption amount as a function of the adsorption potential (${A}_{\textrm{pot}}$) defined as:
\begin{equation}
	A_\text{pot} = -RT\ln\frac{p}{p_\text{s}},
	\label{eq:eq2_1}
\end{equation}
where ${R}$ is the universal gas constant. The differentiation was performed numerically by the approximation of the isotherms using some empirical functions and calculating their derivatives.

\begin{figure} [!b]
\centerline{
\includegraphics[width=0.75\textwidth]{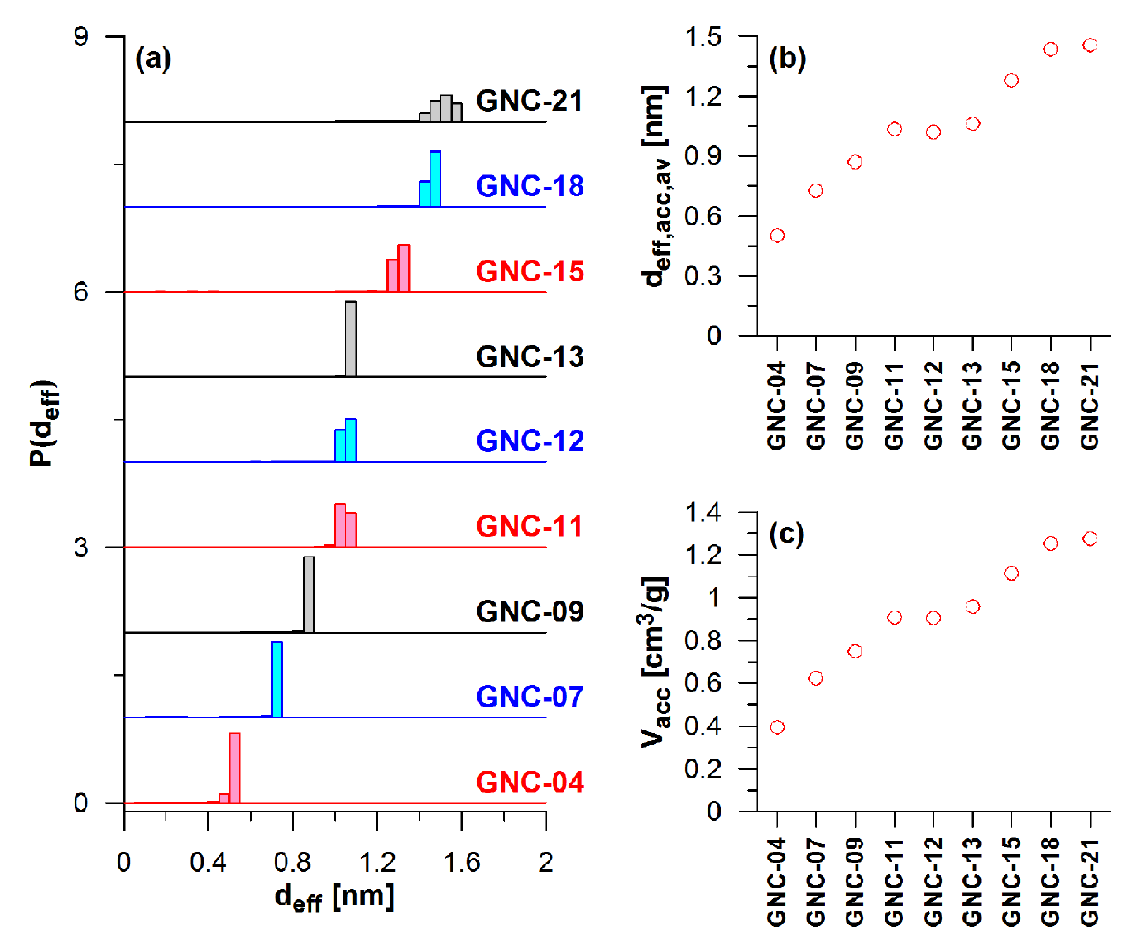}
}
	\caption
	{(Color online) Comparison of (a) pore size histograms for all the studied nanocarbons determined from the BG
	method (the subsequent histograms are shifted by 1 unit from the previous ones), (b) average
	sizes and (c) volume of pores accessible for Ar atoms, respectively.}
	\label{fig:fig_1}
\end{figure}

We also simulated the adsorption and separation of three binary gas mixtures (important from practical point of view): CO$_2$/CH$_4$, CO$_2$/N$_2$, and CH$_4$/N$_2$ on all the studied GNCs. The computations were performed for ${T}=298$~K using the grand canonical Monte Carlo method (GCMC) \cite{Ref91,Ref96}. The simulation scheme was the same as in our previous works \cite{Ref60,Ref62}. Simulations were performed for the total mixture pressure ${p}_{\textrm{tot}}=0.1$~MPa (i.e., atmospheric pressure) and for the following mole fractions of components in the gaseous phase ($y$): 0.0, 0.01, 0.025, 0.05, 0.1, 0.2, 0.3, 0.4, 0.5, 0.6, 0.7, 0.8, 0.9, 0.95, 0.975, 0.99, and 1.0. For each point, the mole fractions of components in the adsorbed phase (${x}$) were calculated from the average numbers of molecules present in the simulation box. The efficiency of the process of separation of mixtures was reflected by the value of equilibrium separation factor (the $1^{\textrm{st}}$ component over the $2^{\textrm{nd}}$ one):
\begin{equation}
	S_{1/2} = \frac{x_1/x_2}{y_1/y_2}.
	\label{eq:eq2_2}
\end{equation}
The adsorbed phase is enriched in the 1$^{\textrm{st}}$ component if ${S}_{{1/2}}>1$.

\section{Results and discussion}

\begin{figure}[!b]
	%\vspace{-2ex}%
\centerline{
\includegraphics[width=0.9\textwidth]{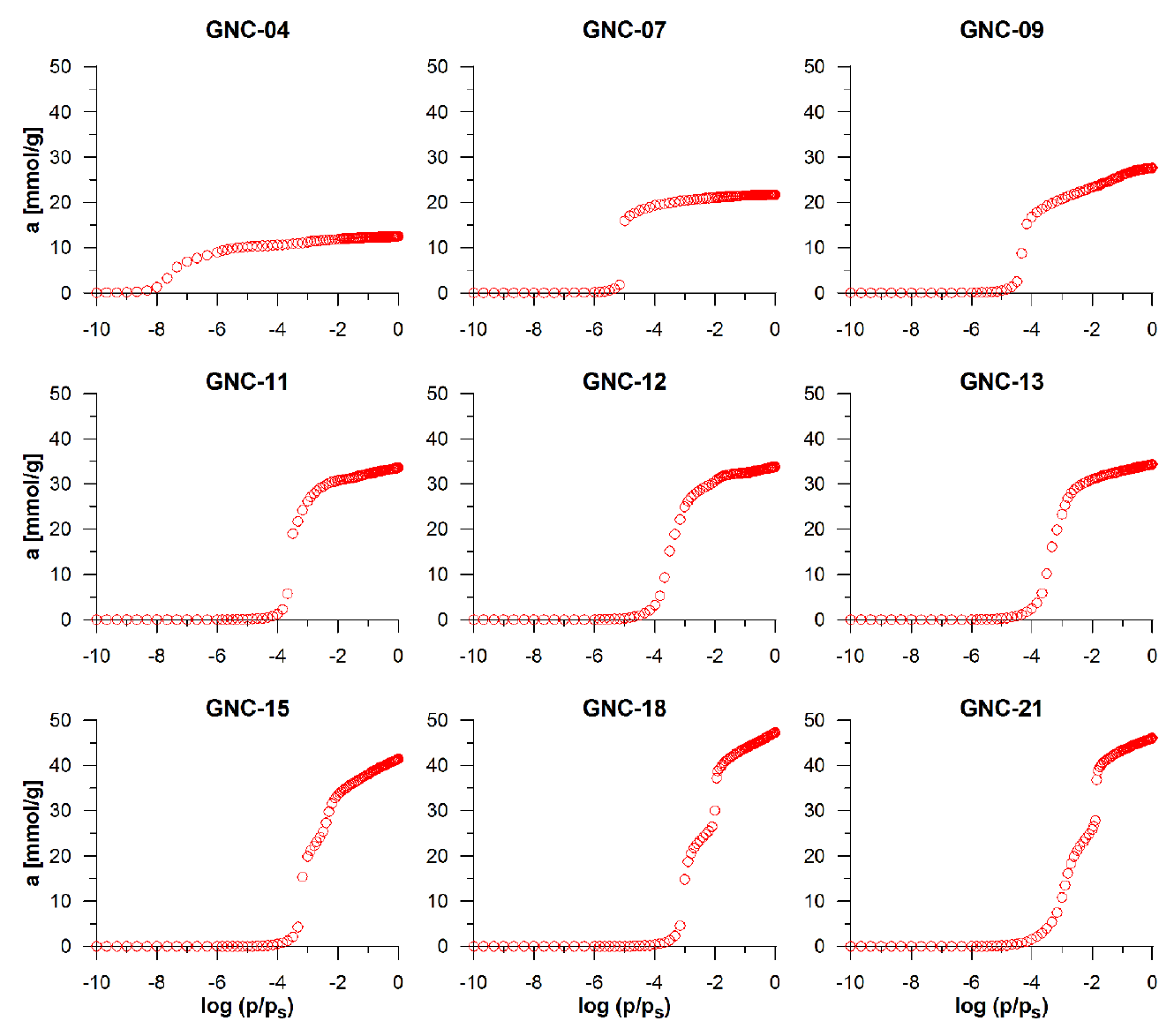}
}
	\caption
	{(Color online) Comparison of Ar adsorption isotherms (\textit{T}~=~87~K) simulated for
	all the considered nanocarbons.}
	\label{fig:fig_2}
\end{figure}

Figure~\ref{fig:fig_1}~(a) collects histograms of effective pore sizes, determined from the application of BG method, for all the studied GNCs. All the structures are strictly microporous, i.e., the diameters of pores do not exceed 2~nm. Generally, the size of dominant pores increases in the considered series (from the GNC-04 up to GNC-21). However, there are two groups of nanocarbons having similar diameters of the main pores: (i) GNC-11, GNC-12 and GNC-13~--- ${d}_{\textrm{eff}}$ around 1~nm and (ii) GNC-18 and GNC-21~--- ${d}_{\textrm{eff}}$ around 1.5~nm. It should be noted that in the case of GNC-21, some amount of pores wider than in GNC-18 is also present. These regularities are reflected by the values of the average pore diameter [figure~\ref{fig:fig_1}~(b)]. The increase in pore diameters is accomplished by the increase in pore volume from ca. 0.4~cm$^{3}$/g for GNC-04 up to ca. 1.3~cm$^{3}$/g for GNC-18 and GNC-21 [figure~\ref{fig:fig_1}~(c)].

\begin{figure}[!t]
	%\vspace{-2ex}%
\centerline{
\includegraphics[width=0.85\textwidth]{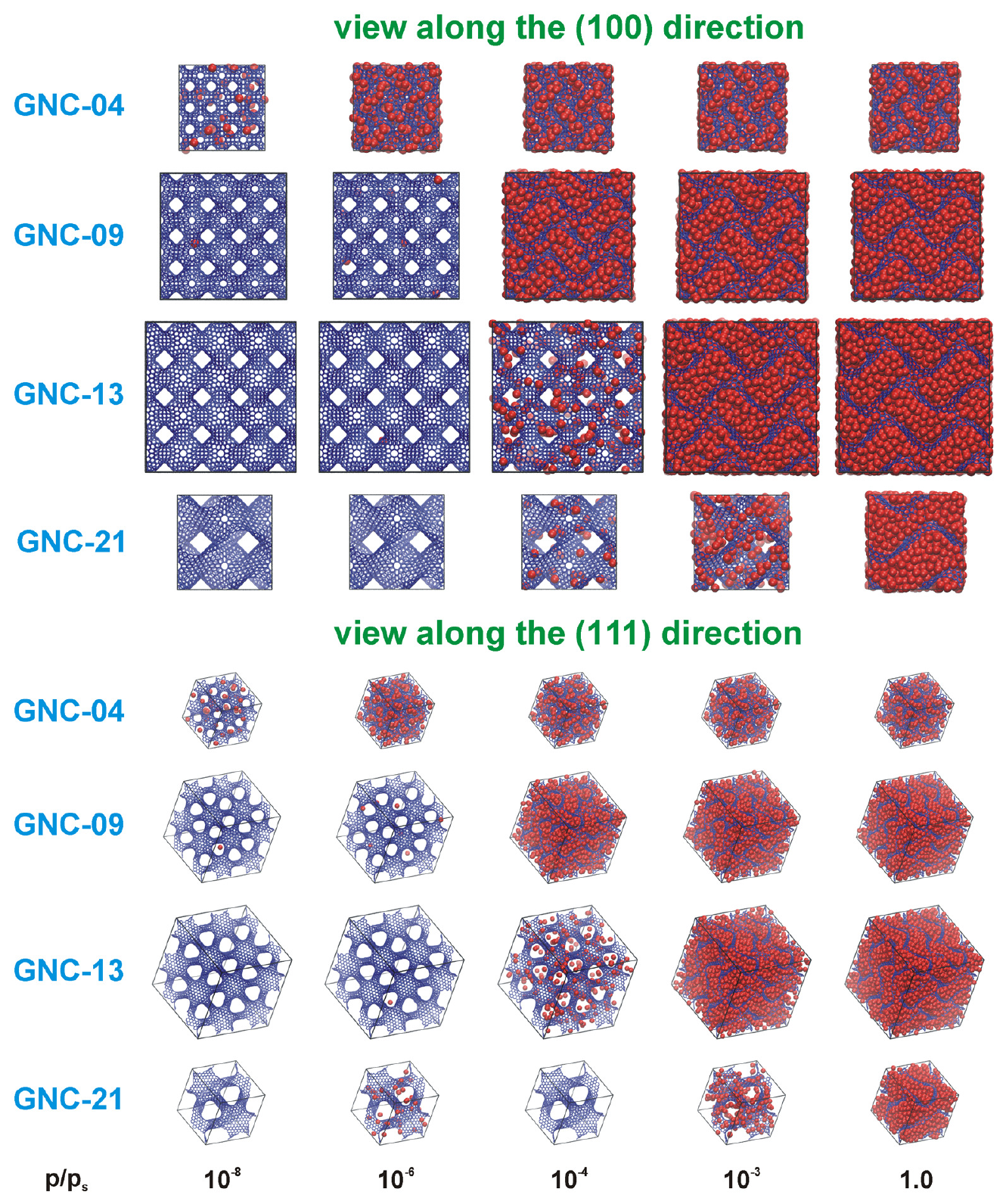}
}
	\caption
	{(Color online) Equilibrium argon configurations for selected nanocarbons and selected values of
	relative pressure (the frames reflect the size of simulation boxes, all the structures are in
	the same scale). It should be noted that this figure was created using the VMD program \cite{Ref97}.}
	\label{fig:fig_3}
\end{figure}

Figure~\ref{fig:fig_2} shows the comparison of Ar adsorption isotherms simulated for all the studied GNCs. The changes observed in the shape of isotherms reflect the differences in the properties of nanocarbons. Adsorption capacity varies from ca. 12~mmol/g for GNC-04 up to ca. 45~mmol/g for GNC-18 and GNC-21. These changes correspond to the differences in the pore volume [figure~\ref{fig:fig_1}~(c)]. At the same time, the shift of the pore filling pressure toward higher values is observed. The pores of the GNC-04 structure are filled in the relative pressure range $10^{{-8}}-10^{{-6}}$. However, the total filling of GNC-18 and GNC-21 occurs for similar values of relative pressure (around $10^{{-2}}$). The middle carbons of the series (i.e., GNC-11, GNC-12 and GNC-13) are filled in the similar range of relative pressure (${p/p}_{\textrm{s}} > 10^{{-4}}$). The differences in the pore filling are also clearly seen on equilibrium snapshots of Ar configurations in the simulation boxes shown in figure~\ref{fig:fig_3}. These regularities are related to the differences in diameters of dominant pores present in the individual GNCs [figure~\ref{fig:fig_1}~(a)]. Finally, one can observe that in the case of initial structures (up to GNC-13), the pore filling is a single-step process. However, the pores of GNC-15, GNC-18 and GNC-21 are filled in two steps. This is also caused by the rise in pore sizes. For pores wider than 1~nm, in the first step a monolayer is formed and next the remaining free volume is filled.

\begin{figure}[!t]
	%\vspace{-2ex}%
\centerline{
\includegraphics[width=0.92\textwidth]{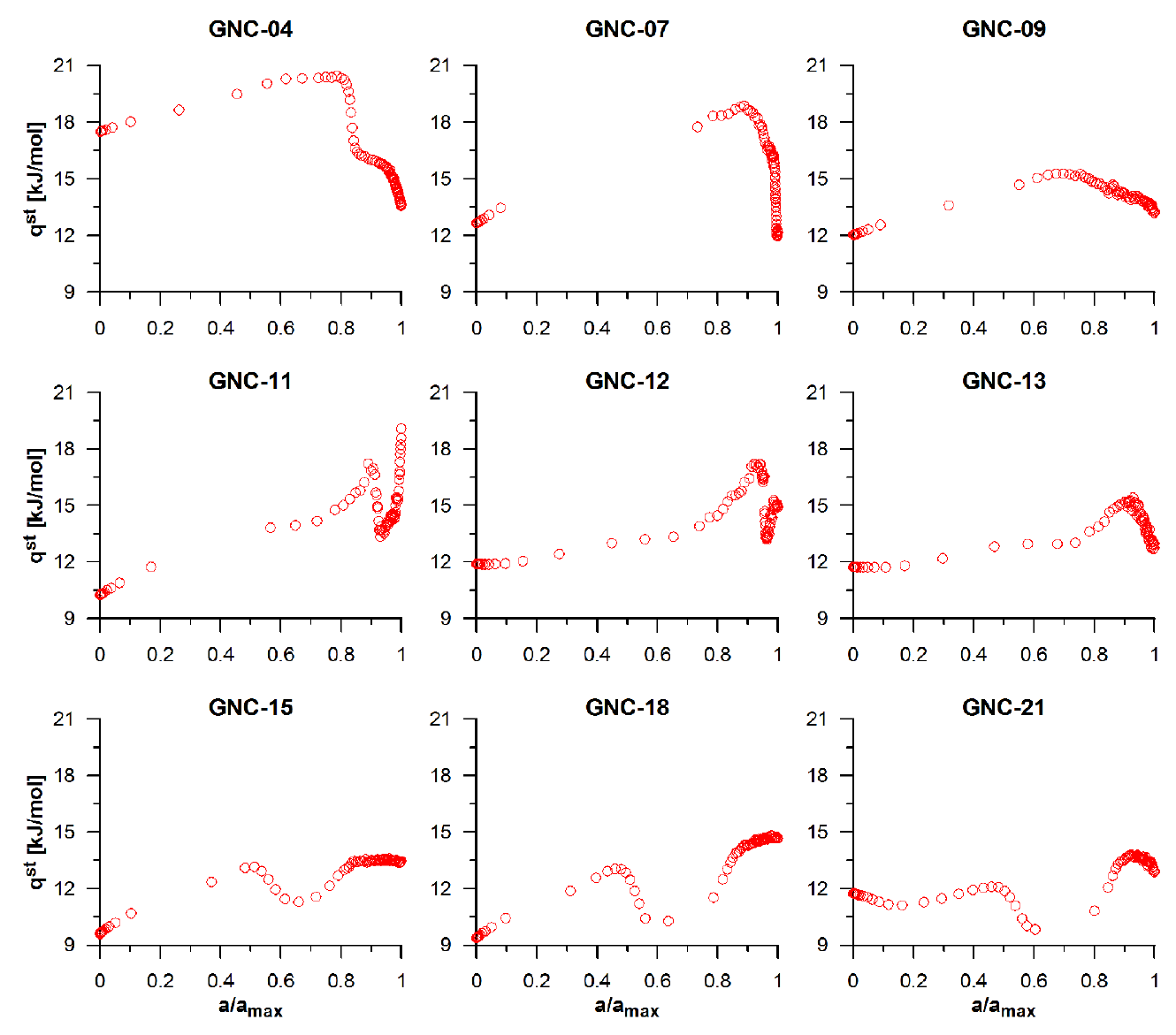}
}
	\caption
	{(Color online) Comparison of isosteric enthalpy of Ar adsorption [the data related to
	figure~\ref{fig:fig_2}; for clarity adsorption amounts are normalised by
	the maximum observed value (${a}_{\textrm{max}}$)].}
	\label{fig:fig_4}
\end{figure}

Figure~\ref{fig:fig_4} shows the plots of isosteric enthalpy of adsorption related to the isotherms shown in figure~\ref{fig:fig_2}. For low loadings, the ${q}^{\textrm{st}}$ values increase as the adsorption amount rises. In this range, Ar atoms are adsorbed mainly on high-energetic centres. The increasing adsorption amount causes other Ar atoms to appear in the vicinity of the initially adsorbed ones. This rises the contribution of fluid-fluid interactions to the total energy of a system. Here, the only exception is GNC-21 structure. There is observed a decrease in ${q}^{\textrm{st}}$ for relative adsorption up to ca. 0.1. This system has probably got a heterogeneous surface. Consequently, the subsequent Ar atoms are adsorbed on centres having lower energy and this reduces the effects of the increase in lateral Ar-Ar interactions. In the intermediate range of adsorption, the enthalpy rises for all the structures until the entire adsorbent surface is covered. Next, the values of ${q}^{\textrm{st}}$ decrease since Ar is adsorbed at the places more distant from the surface and this is connected with lower solid-fluid contributions. In the case of the structures having the widest pores (especially GNC-15, GNC-18 and GNC-21), the second peak on ${q}^{\textrm{st}}$ is also observed. This peak is related to the total filling of pores.

%\begin{wrapfigure}{i}{0.5\textwidth}
\begin{figure}[!t]
	%\vspace{-2ex}%
\centerline{
\includegraphics[width=0.42\textwidth]{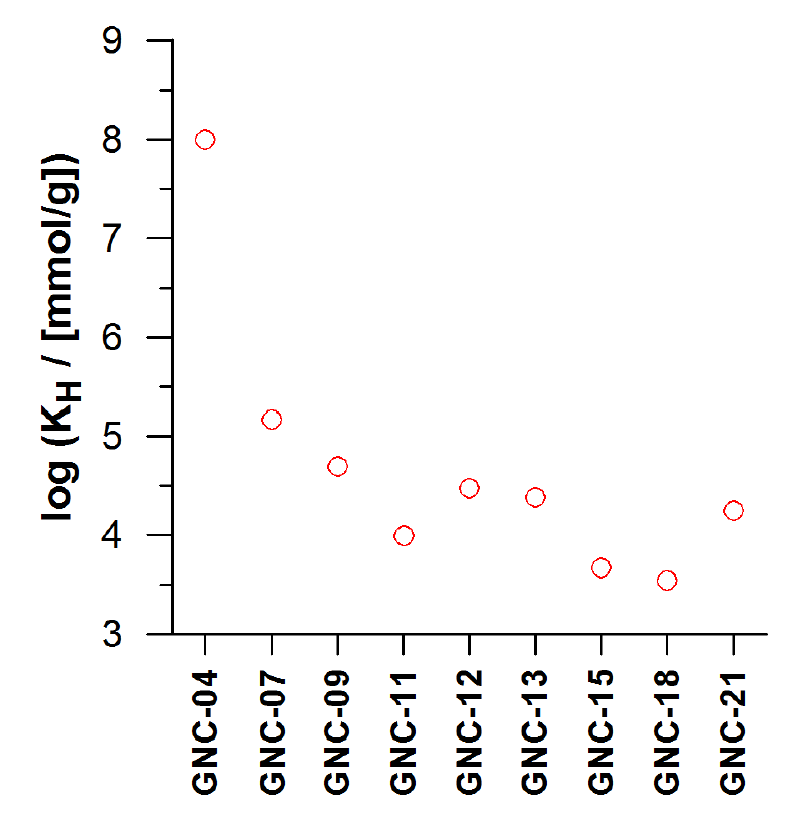}
}
	\caption
	{(Color online) Comparison of Henry’s constants related to the Ar adsorption
	isotherms presented on figure~\ref{fig:fig_2}.}
	\label{fig:fig_5}
\end{figure}
%\end{wrapfigure}
%
\begin{figure}[!b]
	%\vspace{-2ex}%
\centerline{
\includegraphics[width=0.9\textwidth]{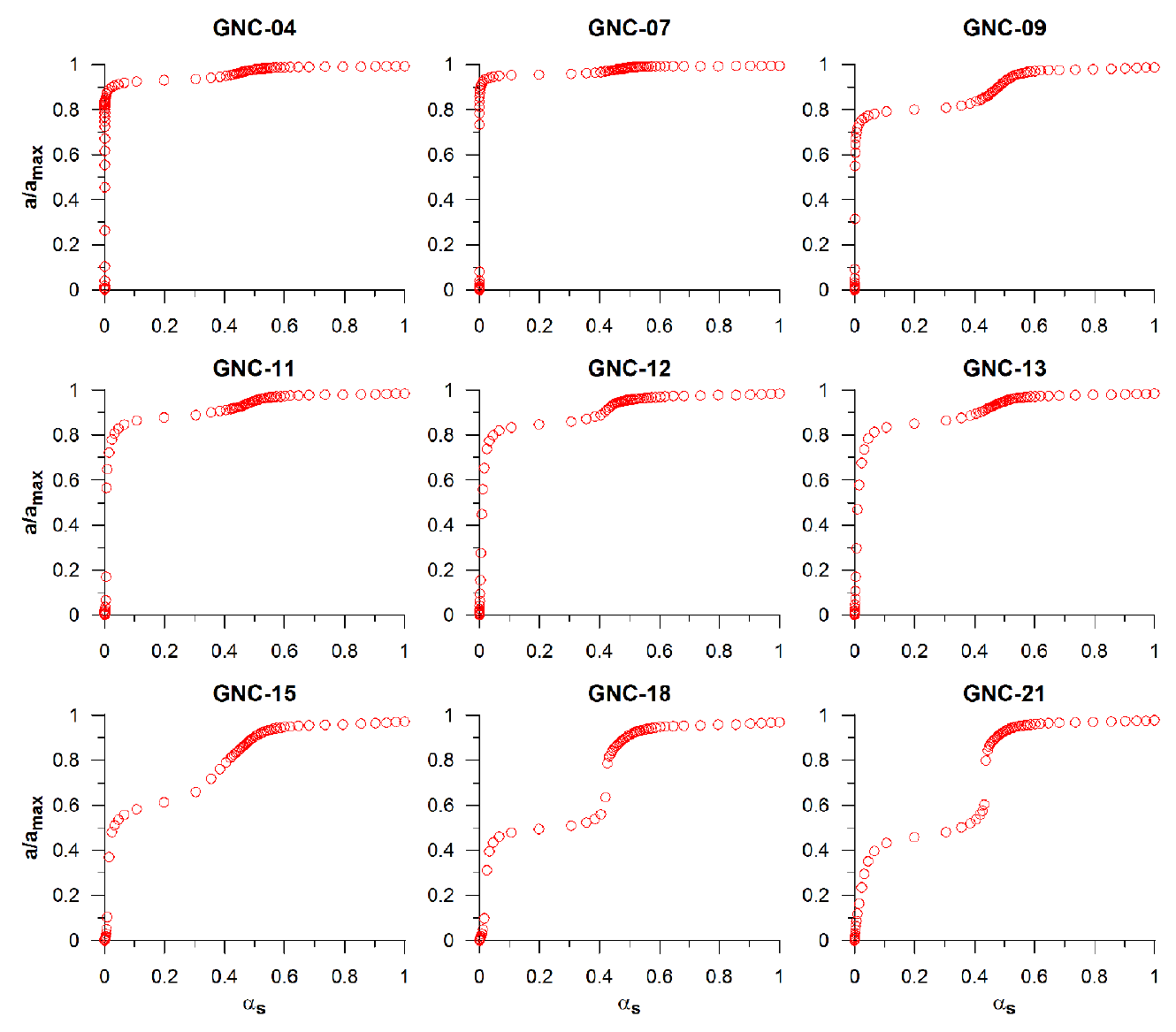}
}
	\caption
	{(Color online) Comparison of $\alpha_{\textrm{s}}$-plots related to the Ar adsorption
	isotherms presented in figure~\ref{fig:fig_2} [for clarity the adsorption amounts are normalized by the
	maximum observed value (${a}_{\textrm{max}}$)].  The $\alpha_{\textrm{s}}$ is the normalized adsorption on the reference material, i.e., in the ideal slit-like system with the effective pore width equal to 10~nm.}
	\label{fig:fig_6}
\end{figure}

Comparing the enthalpy at zero coverage for all the systems, one can observe that GNC-04 has a remarkably higher value of this parameter (ca. 17.5~kJ/mol) than the other systems. This is connected with the presence of the narrowest pores having high adsorption energy. The other structures may be divided into two groups. The first one includes GNC-07, GNC-09, GNC-12, GNC-13 and GNC-21 carbons. In this case, the enthalpy at zero coverage is close to 12 kJ/mol. For the second group (GNC-11, GNC-15 and GNC-18), this enthalpy value is in the range 9--10~kJ/mol. This may suggest some similarities in the surface nature of the group members (for example curvature, which is the main factor determining the energy of adsorption on the surface). The above-described differences in the energy of interactions with the surface of adsorbents fully correspond to the variation of Henry’s constants shown in figure~\ref{fig:fig_5}. This is not surprising since solid-fluid interactions are the main factor affecting the shape of the isotherm in the low pressure range.
Hence, the value of ${K}_{\textrm{H}}$ for GNC-04 system is at least 1000 times greater than for the other ones. For three of the remaining structures (i.e., GNC-11, GNC-15 and GNC-18), lower values of ${K}_{\textrm{H}}$ ($< 10^{{4}}$~mmol/g) are recorded. The same carbons have the lowest enthalpy of adsorption at zero coverage.

Figure~\ref{fig:fig_6} presents comparison of $\alpha_{\textrm{s}}$-plots related to the Ar adsorption isotherms. One can see that the adsorption process is dominated by a FS swings (GNC-04 and GNC-07) and the FS-CS swings (remaining structures) \cite{Ref98}. It can be noticed that with the rise in the pore diameter, the combination of primary and secondary micropore filling mechanism occurs. The boundary between those mechanisms is located for the structures with pore diameters around ca. 0.8~nm.  It is also interesting that the range on $\alpha_{\textrm{s}}$-plots connecting FS and CS swings is not linear as it is observed for the case of slit-like carbon micropores. This can be caused by the surface curvature.

\begin{figure}[!t]
	%\vspace{-2ex}%
\centerline{
\includegraphics[width=0.84\textwidth]{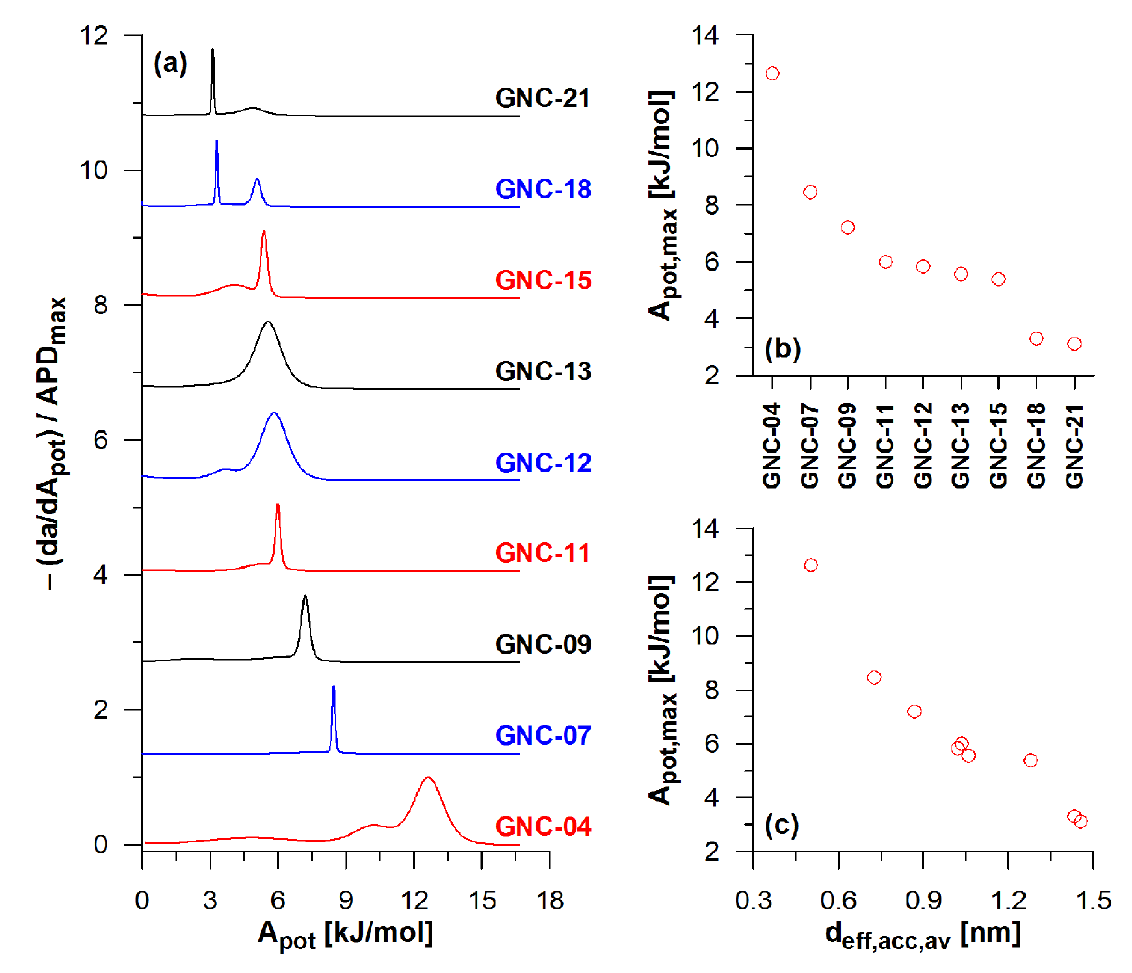}
}
	\caption
	{(Color online) (a) Comparison of APD curves [for clarity, all the curves are normalized by
	the maximum observed value ({APD}$_{\textrm{max}}$); subsequent curves are shifted by
	1.35 units]. (b) Location of the main peak on the APD curves (${A}_{\textrm{pot,max}}$).
	(c) Correlation between the location of the main peak and the average sizes
	of pores accessible for Ar atoms.}
	\label{fig:fig_7}
\end{figure}

Figure~\ref{fig:fig_7}~(a) compares APD curves for all the systems studied. The presented data are complementary to adsorption isotherms shown in figure~\ref{fig:fig_2}. On all the APD curves, at least one (dominant) peak is observed. It corresponds to the pore filling. Its location [${A}_{\textrm{pot,max}}$, figure~\ref{fig:fig_7}~(b)] is related to the pressure of the pore filling according to equation~\ref{eq:eq2_1}. Hence, this parameter may be correlated with the size of pores~--- see figure~\ref{fig:fig_7}~(c). This figure quantitatively confirms the above-described qualitative differences in the pore filling process. The width of the main peak also provides some information on the process. The narrow peak means that condensation occurs in a narrow range of relative pressure. By contrast, a wide peak denotes a wide condensation range. For example, the pore filling in GNC-04 system occurs, as mentioned above, for $10^{{-8}} < {p/p}_{\textrm{s}} < 10^{{-6}}$ and this is reflected by a wide peak with the maximum located at ca. 12.6~kJ/mol. For this system, the other two peaks are also observed (the third one with the maximum at ca. 45~kJ/mol is very broad). These peaks reflect the other sub-steps of the Ar density rise in pores of this structure. Similar interpretation also concerns the additional peaks observed for GNC-11 and GNC-12. In the case of GNC-18 and GNC-21 structures, the observed second peak is related to the above mentioned monolayer formation. A slightly different scenario occurs for GNC-15 carbon. Here, the dominant peak is connected with the monolayer formation and the second low (also wide) peak reflects the filling of the remaining pore volume. This fact explains why the location of the main peak for this structure deviates from the distinct trend visible for all the other GNCs in correlation shown in figure~\ref{fig:fig_7}~(c).

\begin{figure}[!t]
	%\vspace{-2ex}%
\centerline{
\includegraphics[width=0.9\textwidth]{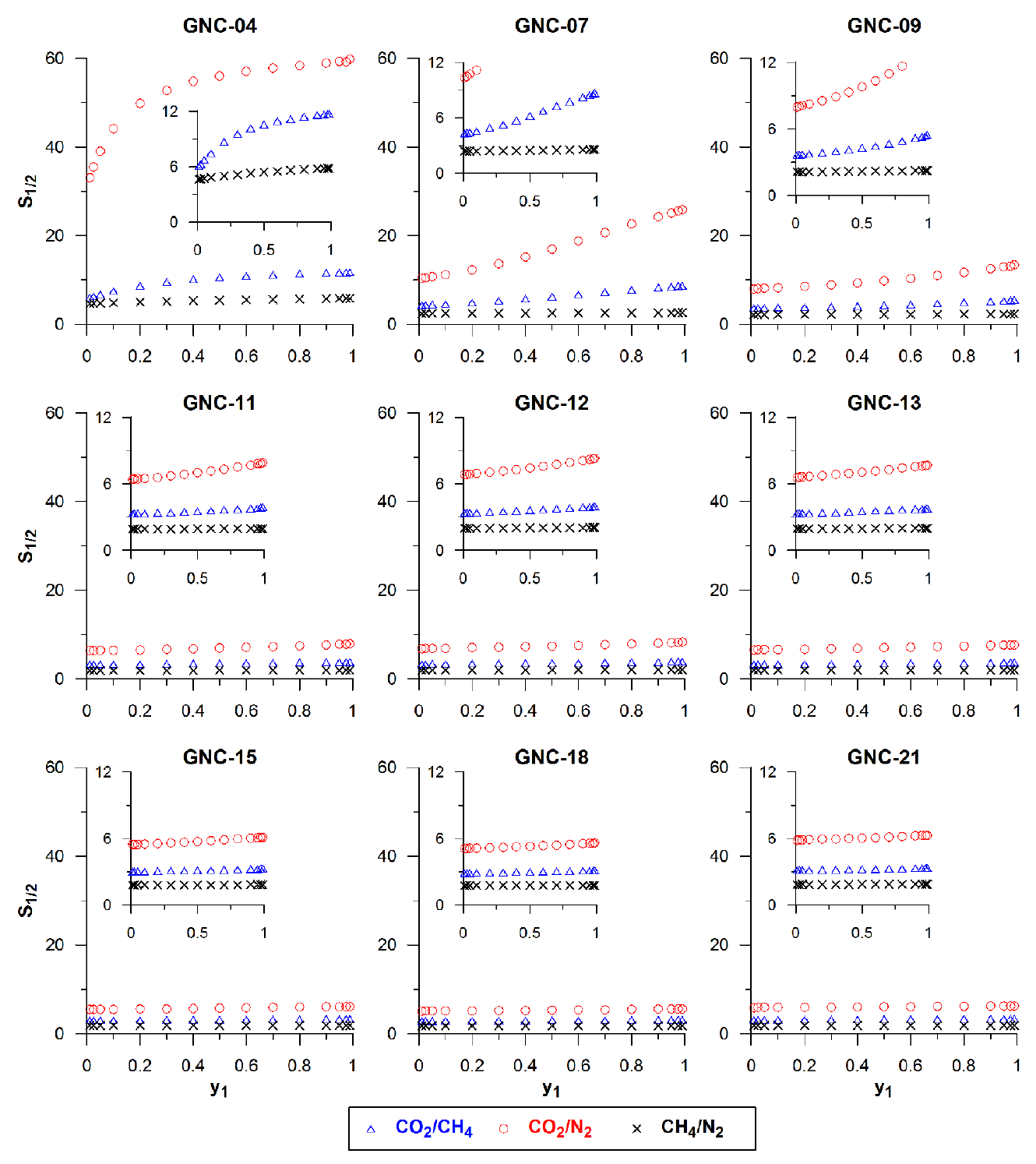}
}
	\caption
	{(Color online) Comparison of equilibrium separation factors
	[${S}_{{1/2}}$, equation~(\ref{eq:eq2_2})] for the adsorption of
	all the mixtures (${T}=298$~K, ${p}_{\textrm{tot}}=0.1$~MPa)
	on all the  nanocarbons studied. The data plotted as the function of
	the 1$^{\textrm{st}}$ component mol fraction in the gaseous phase
	(${y}_{{1}}$, the 1$^{\textrm{st}}$ component is CO$_2$ for
	CO$_2$/CH$_4$ and CO$_2$/N$_2$ mixtures and CH$_4$ for CH$_4$/N$_2$ mixture).}
	\label{fig:fig_8}
\end{figure}

\begin{figure}[!t]
	%\vspace{-2ex}%
\centerline{
\includegraphics[width=0.97\textwidth]{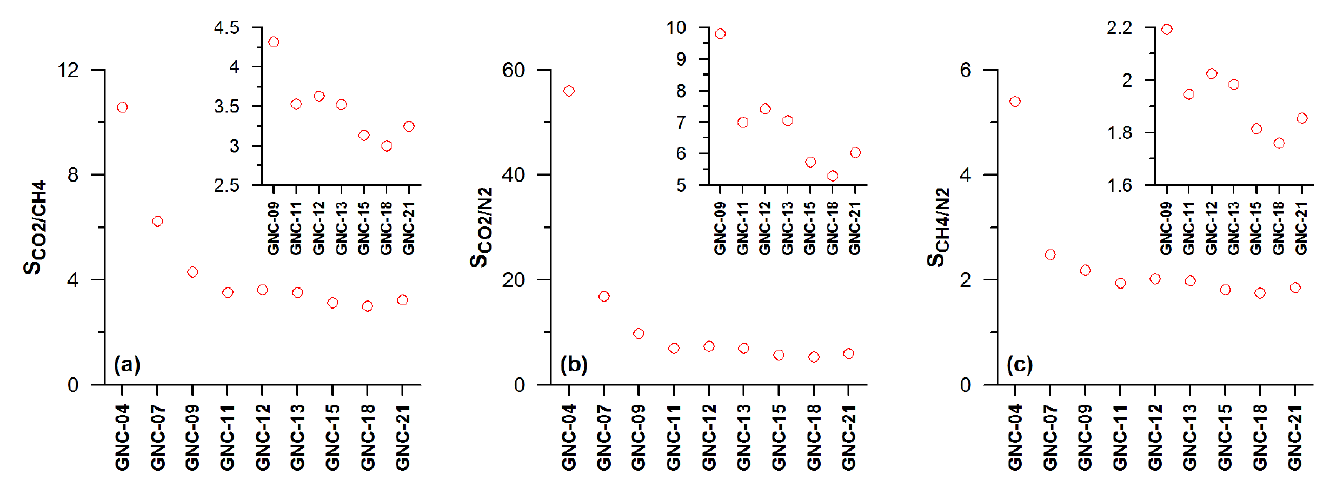}
}
	\caption
	{(Color online) Comparison of equilibrium separation factors for adsorption of
	equimolar \mbox{(${y}=0.5$)} mixtures: (a) CO$_2$/CH$_4$, (b) CO$_2$/N$_2$
	and (c) CH$_4$/N$_2$ (${p}_{\textrm{tot}}=0.1$~MPa,
	${T}=298$~K)~--- see figure~\ref{fig:fig_8}.}
	\label{fig:fig_9}
\end{figure}

\begin{figure}[!b]
	%\vspace{-2ex}%
\centerline{
\includegraphics[width=0.81\textwidth]{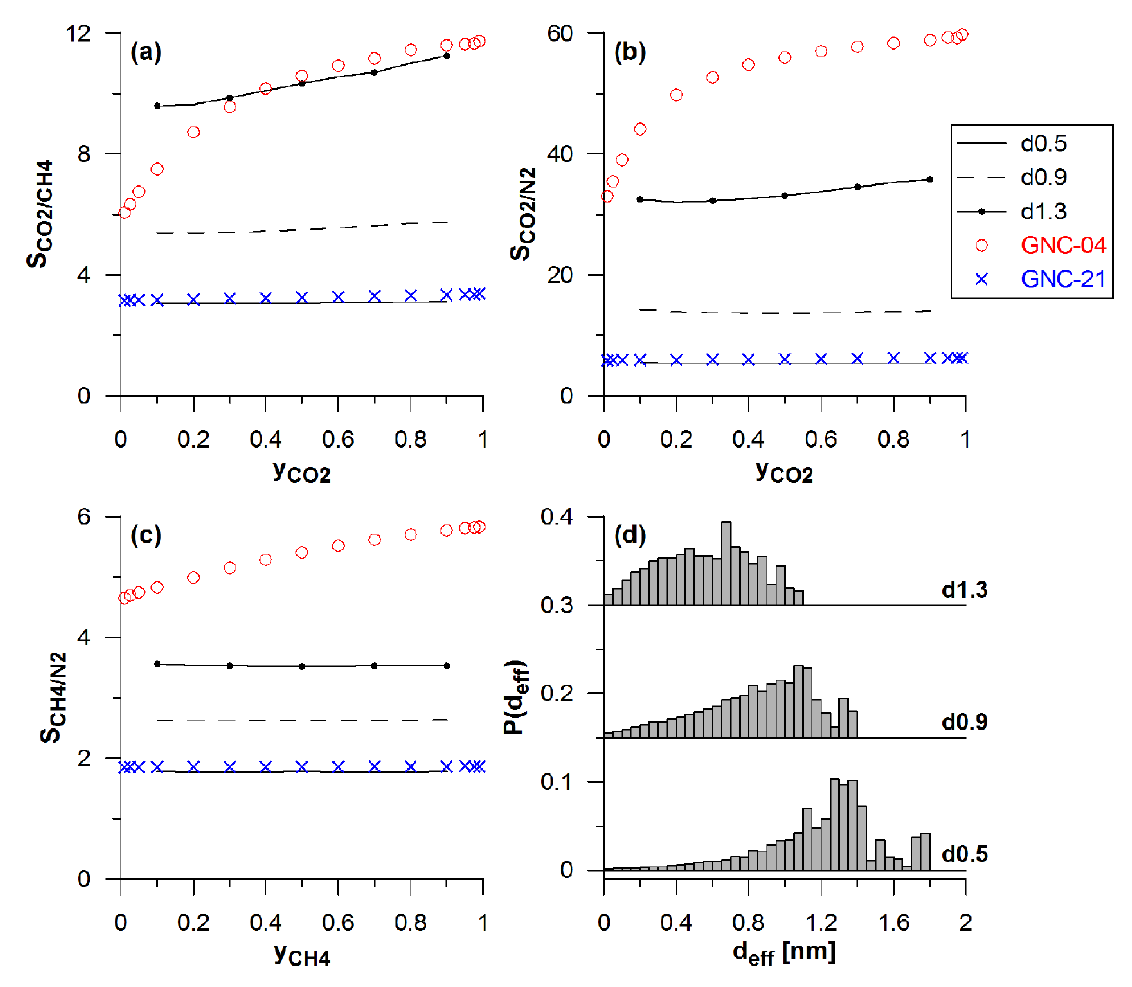}
}
	\caption
	{(Color online) Comparison of equilibrium separation factors
	[(a) CO$_2$/CH$_4$, (b) CO$_2$/N$_2$ and (c) CH$_4$/N$_2$] for mixture adsorption
	(${p}_{\textrm{tot}}=0.1$~MPa, ${T}=298$~K) on GNC-04 and GNC-21 nanocrabons
	and on three virtual porous carbons (VPCs): d0.5, d0.9 and d1.3 \cite{Ref88}
	(the data for VPCs taken from our previous paper \cite{Ref60}). In addition,
	panel (d) presents the pore size histograms for the presented
	VPCs \cite{Ref88} (the subsequent histograms are shifted by
	0.15 units from the previous ones).}
	\label{fig:fig_10}
\end{figure}

Figure~\ref{fig:fig_8} presents a comparison of equilibrium separation factors for adsorption of all three studied mixtures for different compositions of gaseous phase. In addition, figure~\ref{fig:fig_9} directly compares the efficiency of separation of equimolar mixtures for all the studied systems. The separation is a consequence of differences in the adsorption of mixture components. Since the critical temperature for the studied gases decreases significantly in the sequence CO$_2$~>~CH$_4$~>~N$_2$, the adsorption affinity decreases in the same sequence (for the subsequent gases, the process occurs for increasing value of reduced temperature). Consequently, for the given GNC, the highest equilibrium separation factor is observed for CO$_2$/N$_2$ and the smallest one for CH$_4$/N$_2$ mixture. The qualitative differences in separation efficiency between the structures studied are the same for all the mixtures studied. The highest values of equilibrium separation factor (remarkably higher than for the other systems) are observed for GNC-04 carbon. This is connected with the presence of the narrowest pores. Among the other nanocarbons,  GNC-07 and GNC-09 also exhibit higher values of ${S}_{{1/2}}$. However, they are lower than in the case of GNC-04 since these structures have wider pores. The efficiency of separation for the next GNCs is similar. Nevertheless, some small differences are also noticeable (see insets in figure~\ref{fig:fig_9}). The GNC-11, GNC-15 and GNC-18 structures are less efficient in comparison with the adjacent carbons in the series. These regularities are analogous to that observed for Henry’s constants shown in figure~\ref{fig:fig_5} and discussed above. These facts suggest that in the case of GNCs with pores wider than ca. 1~nm, the main factor affecting the efficiency of the separation process is the energetics of fluid interaction with the curved surface of the  nanocarbons studied.

Finally, figure~\ref{fig:fig_10} compares the equilibrium separation factors (all three mixtures studied) for GNC-04 and GNC-21 structures and for three virtual porous carbons (VPCs) described in detail previously \cite{Ref88} and having different porosity~--- see figure~\ref{fig:fig_10}~(d). As one can see, the GNC-21 structure exhibits a separation efficiency similar to the d0.5 carbon. The main pores of both adsorbents have a similar width. However, this VPC has also some narrower micropores which presence probably positively affects the separation efficiency. Such micropores are absent in the case of GNC-21 and, nonetheless, this nanocarbon exhibits similar values of ${S}_{{1/2}}$. This is the consequence of the adsorption energetics on a curved surface of this structure. On the contrary, the GNC-04 nanocarbon exhibits the efficiency of the CO$_2$/CH$_4$ mixture separation similar to the d1.3 carbon. This VPC has micropores distributed in the range up to ca. 1~nm [figure~\ref{fig:fig_10}~(d)]. A large part of them has diameters  similar to or lower than the GNC-04 structure. In the case of CO$_2$/N$_2$ and CH$_4$/N$_2$ mixtures, the values of equilibrium separation factors for GNC-04 are higher than for  d1.3 carbon. This comparison (especially for CO$_2$/CH$_4$ mixture~--- similar efficiency for both adsorbents and for CH$_4$/N$_2$ mixture~--- higher efficiency for GNC-04) may suggest that a regularly curved surface of gyroidal carbons exhibits higher affinity to CH$_4$ molecules than a heterogeneous surface of activated carbons.

Summing up, the GNCs studied may be considered as potential adsorbents for gas mixture separation. The efficiency of this process is similar to or higher than for activated carbons with similar diameters of pores. The GNC-04 or similar structures seem to be  quite promising materials for this purpose since this nanocarbon contains narrow and quite uniform pores (ca. 0.5~nm).

\section{Conclusions}

Adsorption and separation properties of GNCs~--- a new class of exotic nanocarbon materials, are studied for the first time using computer simulation technique. All the structures studied are strictly microporous. The mechanisms of Ar adsorption are described basing on the analysis of adsorption isotherms, enthalpy plots, the values of Henry’s constants, $\alpha_{\textrm{s}}$ and adsorption potential distribution plots. Below the pore diameters ca. 0.8~nm, primary micropore filling process dominates. For structures possessing larger micropores, primary and secondary micropore filling mechanisms are observed. GNCs may be considered as potential adsorbents for gas mixture separation, having separation efficiency similar to or higher than this for activated carbons with similar diameters of pores.

\section*{Acknowledgements}

The authors acknowledge the use of the computer cluster at Pozna\'{n} Supercomputing and Networking Centre (Pozna\'{n}, Poland) as well as the Information and Communication Technology Centre of the Nicolaus Copernicus University (Toru\'{n}, Poland).

%\clearpage

\ukrainianpart

\title
{
Гіроїдні нанопористі вуглеці~--- адсорбція і особливості розділення, досліджені з використанням \\ комп'ютерного моделювання}
\author{С. Фурманяк\refaddr{label1},
                            П.А. Гауден\refaddr{label1},
                            А.П. Тержик\refaddr{label1},
                            П. Ковальчик\refaddr{label2}}
\addresses{
\addr{label1} Хімічний факультет, Дослідницька група фізикохімії вуглецевих матеріалів, \\ Університет Ніколауса Копернікуса в Торуні, Торунь, Польща
\addr{label2} Школа інженерії та інформаційних технологій,  Університет Мердока, \\ Мердок, Західна Австралія,  6150, Австралія
}

\makeukrtitle

\begin{abstract}
Адсорбція і особливості розділення у гіроїдних нанопористих вуглецях  (GNC), новому класі екзотичних нановуглецевих матеріалів, вперше досліджено, використовуючи метод моделювання Монте Карло з гіперпаралельним темперуванням.
 Пориста структура  GNC моделей оцінюється методом, запропонованим Бхатачарія і Губінсом. Всі досліджені структури  є строго мікропористі. Крім того,  механізми адсорбції Ar описуються на основі  аналізу ізотерм адсорбції, кривих ентальпії, значень сталих Генрі, $\alpha_\textrm{s}$
 та кривих розподілу  потенціалу адсорбції. Зроблено висновок, що при діаметрах пор близьких або менших за 0.8~nm домінує процес заповнення первинних мікропор. Для структур, що мають більші мікропори, спостерігається механізм заповнення первинних та вторинних мікропор. Нарешті, описано властивості розділення CO$_2$/CH$_4$, CO$_2$/N$_2$ і CH$_4$/N$_2$ сумішей у GNC і виконано порівняння з властивостями розділення  моделей Віртуального Пористого Вуглецю.
  Гіроїдні пористі вуглеці можна розглядати як потенційні адсорбенти для розділення газових сумішей, які володіють аналогічною або навіть вищою ефективністю розділення, ніж активований вуглець з подібним діаметром пор.

\keywords гіроїдні нанопористі вуглеці, адсорбція, розділення газових сумішей, моделювання Монте Карло
\end{abstract}

\end{document}